# Structural and electrical characterization of $NbO_2$ thin film vertical devices grown on TiN-coated $SiO_2$/Si substrates

Toyanath Joshi[1], Pavel Borisov[2], and David Lederman[1]

[1]*Department of Physics, University of California Santa Cruz, Santa Cruz, CA 95064, USA*
[2]*Department of Physics, Loughborough University, Loughborough, LE11 3TU, UK*

**We report on the electrical properties of polycrystalline $NbO_2$ thin film vertical devices grown on TiN coated $SiO_2$/Si substrates using pulsed laser deposition. First, we analyzed the thickness and contact size dependences of threshold switching of $NbO_2$ films grown in 10 mTorr Ar/$O_2$ mixed growth pressure, where 25.1%/74.9% of $NbO_2$/$Nb_2O_5$ surface composition content was estimated by *ex-situ* x-ray photoelectron spectroscopy. The leakage current properties in the insulating state were dominated by trap-charge assisted Poole-Frankel conduction mechanism. The threshold switching and self-sustained current oscillatory behavior of films with different $NbO_2$/$Nb_2O_5$ composition ratios was measured and analyzed. The second film was grown in lower (1 mTorr) growth pressure which resulted in a higher (34.2%/65.8%) $NbO_2$/$Nb_2O_5$ film surface composition. The film grown in higher growth pressure demonstrated lower off-state leakage current, faster switching and self-sustained oscillations with higher frequency than the film grown in lower growth pressure.**

## I. Introduction

Transition metal oxides demonstrating metal-to-insulator transitions (MIT), such as $VO_2$[1-3] and $NbO_2$, are widely studied materials for the use as selector elements in resistive random access memory (RRAM) and in current oscillatory devices.[1-14] $VO_2$ has shown changes on the order of $10^5$ in resistivity around $T_{MIT}$ = 340 K.[1] The MIT temperature of $VO_2$, however, is close to room temperature and can be easily reached by heat dissipation while used in electronic devices. $NbO_2$, on the other hand has a much higher MIT temperature ($T_{MIT}$ = 1081 K)[15,16] making this material a robust candidate for current threshold switching devices.[4-14]

Previously reported $NbO_2$ thin film vertical devices have been composed of amorphous $NbO_2$ films.[4-7,10-14] In the case of amorphous films, it is proposed that a crystalline conducting path is formed during the initial electroforming process where crystallization is mainly governed by the nucleation of the amorphous thin film material followed by crystal growth.[17,18] A continuous increase in crystalline area and in crystal grain size upon multiple resistance switching cycles leads to reduced device lifetimes. Also, relatively high OFF-state leakage currents >10 $\mu A$[9,11-14] result in larger power dissipation >10 $\mu W$/bit, assuming an operation voltage of 1 V, and thus a reduction in leakage current would be beneficial. Epitaxial films, on the other hand, have longer current switching times and thus are not deemed suitable for applications in devices.[19,20]

Here we report on the structural and electrical properties of $NbO_2$ polycrystalline thin films grown using PLD on TiN coated $SiO_2$/Si substrates, a platform important for the further development of electronic devices based on NbO$_2$. We pursued three promising routes for optimizing the performance of NbO$_{2-x}$ film devices: (1) varying thickness of the film, (2) varying the size of top electrode metallic contacts, and (3) varying Nb$^{4+}$/Nb$^{5+}$ content ratio.

## II. Experimental details

TiN (11 nm)/SiO$_2$(100nm)/Si wafers, provided by Micron Technology, were used as substrates for NbO$_2$ thin film growth using pulsed laser deposition. A ceramic target of Nb$_2$O$_5$ was prepared from Nb$_2$O$_5$ powder (99.99%, Sigma-Aldrich). The powder was pressed into a pellet and sintered for 72 h in air at 1300 °C. The distance between the target and the substrate was 7.3 cm. The KrF laser energy density at the target was approximately 2 J/cm$^2$ and its pulse repetition rate was 5 Hz. The substrate temperature was kept at 700 °C. A set of films with different thicknesses was grown in 10 mTorr O$_2$/Ar mixed gas atmosphere with 7% O$_2$ mass flow. A 1 mTorr growth pressure with 7 % O$_2$ mass flow was used in order to examine the effects of possible oxygen vacancies.

A four-axis goniometer x-ray diffraction (XRD) system with a Cu K$_\alpha$ rotating anode was used for structural characterization of the deposited films. Out-of-plane (θ-2θ) and grazing incidence x-ray diffraction (GIXRD) scans were performed. Polycrystalline peaks of NbO$_2$ were verified from the powder diffraction file PDF# 00-043-1043.[21] Film thickness and surface roughness analysis were performed on x-ray reflectivity (XRR) data obtained from the same x-ray diffraction system.

The chemical composition of the films was analyzed using x-ray photoelectron spectroscopy (XPS) measured in a PHI 5000 VersaProbe spectrometer from Physical

Electronics. The spectra were calibrated using C 1s peak at 284.8 eV and fitted under assumption of a Shirley-type background. Areas under each curve were compared to obtain molar ratios between different Nb oxides present in the films.

The current-voltage (*I-V*) characteristics were measured using probe station tungsten tips (Picoprobe) from GGB Industries Inc. which were pressed directly on top of the film. The nominal lateral size of the probe station tip used was ~2 μm. The voltage was applied across the thin film at ramp rates of 0.1 V/sec and the output current was measured across an attached serial test resistance of $R_s$ = 51 Ω. For pulsed *I-V* measurements, short triangular ramp pulses were applied to the sample and the output current was measured using digital oscilloscope.

Contact-size dependent *I-V* measurements were performed with a voltage loop frequency of 1 Hz using an atomic force microscopy instrument from Asylum equipped with conducting platinum coated Si cantilever tips. Top circular Pt electrodes were sputtered to 50 nm thickness with diameters ranging from 6.5 μm to 0.1 μm using electron-beam lithography.

### III. Experimental results and discussion

#### A. Structural and chemical analysis

Figure 1 shows θ-2θ and grazing incidence x-ray diffraction scans of two of the $NbO_2$ films grown on $TiN/SiO_2/Si$ substrates in 1 and 10 mTorr $Ar+O_2$ (~7% $O_2$) pressures and having thicknesses 43.7 and 33.2 nm, respectively. The presence of $SiO_2$ peaks in the 10 mTorr sample are explained by $SiO_2$ crystallization at the growth substrate temperature in an oxygen atmosphere. Note the absence of $SiO_2$ peaks in the film grown in 1 mTorr pressure due to the lower $O_2$ growth pressure. No peaks related to $Nb_2O_5$ or NbO phases were observed.

Figure 2 shows x-ray reflectivity curves for the samples with thickness ranging from 13.7 to 76.4 nm grown in to 10 mTorr and a 43.7 nm film grown in 1 mTorr (blue open circles symbol). Fitting (solid curves) to an optical model was performed using GenX software.[22] The bottom TiN layer had a thickness of 11 nm with an interface roughness at the TiN/$NbO_2$ interface of ~0.4 nm. The maximum surface roughness was approximately 1 nm for the 76.4 nm thick film. The best fit was achieved assuming an additional top layer with approximately 2.7 nm thick which is most likely a $Nb_2O_5$ layer formed by oxidation of $NbO_2$ after exposure to the atmosphere. The inset in Fig. 2 shows a cross-sectional SEM image of 76.4 nm film. All four layers Si, $SiO_2$, TiN and $NbO_2$ are visible with the corresponding thicknesses agreeing with the nominal values. It is difficult to unequivocally ascertain whether the top $Nb_2O_5$ layer is visible in the SEM image.

The molar content of $NbO_2$ and $Nb_2O_5$ phases was obtained using XPS spectra taken at the 3d core Nb level. XPS spectra measured on the films grown in 1 and 10 mTorr total growth pressures are shown in Fig. 3(a) and 3(b), respectively. The deconvoluted peak fits for the film grown in 10 mTorr pressure show $3d_{3/2}$ and $3d_{5/2}$ peaks from the $Nb^{4+}$ state at 205.6 and 208.3 eV, respectively, in agreement with previous reports.[19,23] Large peaks at 207.1 and 209.8 eV were observed corresponding to $Nb^{5+}$, probably due the oxidation of the film surface to $Nb_2O_5$ after exposure to the atmosphere. The shaded areas mark the peaks related to $Nb^{4+}$ which correspond to the $NbO_2$ phase. The approximate content of $NbO_2$ and $Nb_2O_5$, was calculated using the area covered by $3d_{5/2}$ level peaks, and yielded the values of 34.2%/65.8% and 25.1%/74.9% for the films grown in 1 and 10 mTorr growth pressures, respectively. Peaks related to $Nb_2O_5$ were not visible in XRD or GIXRD spectra in Fig. 1. This suggests that either the $Nb_2O_5$ top layer was amorphous or at least polycrystalline. Other authors have reported the estimated thickness of the top $Nb_2O_5$ layer to be ≈2 nm.[24] The exact estimation of the oxygen stoichiometry, however, is difficult via *ex-situ* measurements due to the possibility of the film surface oxidation after exposure to the air.

### B. Film thickness effect on current threshold switching

First, we analyzed *I-V* characteristics measured on $NbO_2$ films [Fig. 4(a)] with film thicknesses of 13.7, 33.2 and 76.4 nm grown in 10 mTorr total $Ar+O_2$ pressure. The measurements were taken using the probe station tungsten tip as a top electrode and TiN layer as a bottom contact. Clear threshold switching in the output current was observed and the threshold switching voltage ($V_{th}$) depended linearly on the film thickness [Fig. 4(b)]. The linear dependence of $V_{th}$ indicated that the ratio between the threshold voltage and the thickness, i.e. the threshold electric field, was constant and therefore the bulk portion of the film took part in the switching, rather than the metal-oxide interface. This contrasts with some other materials, for example $Fe_2O_3$,[25] which have a resistive switching behavior that is independent of the film thickness because the metal-oxide interface dominates the switching. The current measured at constant voltage [0.9 V in Fig. 4(b)], on the other hand, showed an exponential increase with decreasing film thickness. As reported by Meyer et al.,[26] the mobility (or the drift velocity) of ionic current carriers increases exponentially with increasing applied electric field (i.e. with decreasing film thickness, if the voltage is constant), causing an exponential increase in the current. Also, defects at the film-substrate interface contribute more to the conductivity in thinner films.

Low voltage ($V < V_{th}$) regions of *I - V* curves of the samples with different thicknesses were fitted to ohmic ($I \sim V$), space-charge-limited ($I \sim V^2$), Schottky [$I \sim \exp(\beta V^{\frac{1}{2}})$], and Poole-Frankel conduction [$I \sim V\exp(\beta V^{1/2})$], T = constant] models (see Fig. 4c for 33.2 nm film). For all three thicknesses, the P-F model[27,28] resulted in the best fits which suggests that the high resistance state conduction was due to field-assisted thermal emission of carriers from Coulomb trap centers.[10] Several other authors also have claimed that P-F conduction dominates in crystalline[10] as well as in amorphous films[29-31] of $NbO_x$.

### C. Top contact size effect on current threshold switching

The dependence of the *I-V* characteristics on contact size was further investigated using the conducting probe AFM technique. For these measurements, the current compliance was set to 10 µA. *I-V* curves obtained from the 10 mTorr growth pressure sample using top contact diameters $d$ ranging from 0.1 µm to 6.5 µm are shown in Fig. 5(a). Leakage currents measured at $V = 0.3$ V as a function of contact area $A$ are shown in the Fig. 5(b). The data are fit well on a log-log scale to two power law segments of the form $\ln(I) = n\ln(A) + C$, where $n$ is the exponent in the power law, $A$ is the contact area, and $C$ is a constant. For larger area contacts with $A > 2.8 \times 10^{-9}$ cm², $n_1$ = 0.82 ± 0.06, or *I* ~ $d^2$, whereas for smaller area contacts with $A \leq 2.8 \times 10^{-9}$ cm², $n_2$ = 0.51 ± 0.02, or *I* ~ *d*. Thus, below the critical area of $A = 2.8 \times 10^{-9}$ cm² [$d$ = 0.6 µm, the dotted line in Fig. 5(b)], the current is proportional to the diameter rather than the area of the contact. Similar studies on $NbO_2/TiO_x/NbO_2$[32] trilayers have revealed that the current depends on the electrode area ($I \propto A$), which implies that the number of carrier traps per unit area, which are responsible for the oxide conductivity, is approximately constant. On the other hand, if $I \propto A^n$ with $n < 1$ (e.g., $n = 0.3$ in Ref. 32 for a low resistance state), the current is not uniform across the contact area. In our case, $n \approx 0.5$, which may indicate that electrode edges are the regions which dominate low-field conduction, in which case the current is proportional to the contact's circumference and therefore $I \propto d \propto A^{1/2}$. A transformation to $n \approx 1$ for diameters $d > 0.6$ μm then corresponds to the diminished influence of the contact edge in larger electrodes. Other reports[12,32] consisted of modifying the bottom electrodes, and so they could have encountered an additional effect due to different film growth mode in the resistive switching layer.

In any case, the dependence of current on contact size is of technological importance. In Si-based FinFET devices, for example, reduction in device dimension results in increased leakage current due to the quantum tunneling of the electrons between the gates. In the case of the $NbO_2$ device, however, lowering the lateral device dimension not only increases the device density, but also lowers the leakage current. Using the smallest contact size (i.e., AFM tip), the leakage current measured at $V_{th} \approx 1.6$ V was $I_{th}$ ~$10^{-6}$ A which is an order of magnitude lower than reported elsewhere to date.[5,11-13]

### D. Effects of $Nb^{5+}/Nb^{4+}$ content

We also investigated the effect of $O_2$/Ar total pressure on the *I-V* characteristics. *I-V* characteristics measured on two films grown in 1 and 10 mTorr growth pressure [Fig. 6(a)] were compared using the top contact with a diameter of 0.3 μm. For currents measured at the same voltage, for example at 0.2 V, the film grown in lower growth pressure (1 mTorr) had a current that was an order of magnitude higher than the one measured on the film grown in higher growth pressure (10 mTorr). The larger content of insulating $Nb_2O_5$ was likely responsible for lower leakage current in the film grown in a higher oxygen pressure. On the other hand, a larger defect density, such as O-vacancies, might be responsible for increased leakage current for the film grown in a lower growth pressure. Partial overlap of Coulomb potential wells with increased trap density helps lower the barrier height in the P-F model, thus increasing the conductivity. The bulk film region might have a small amount of conducting NbO or metallic Nb which might be the origin of larger leakage current. It is not possible to estimate exact bulk film stoichiometry using ex-situ XPS measurements because of the surface sensitive nature of XPS. NbO or metallic Nb on the surface of the film would probably be easily oxidized to $NbO_2$ or $Nb_2O_5$ upon exposure to the atmosphere.

Our previous study of $NbO_2$ lateral devices[19,20] demonstrated that the current switching was caused by Joule heating from the current flowing through the devices. Prior to observing a threshold switching, a forming pulse was needed, requiring the application of larger electric field magnitudes than those used during normal operation. It is believed that during an electroforming step, an accumulation of defects, such as oxygen vacancies, takes place which creates a conductive path for the current. Our measurements indicate that the forming voltage also depends on the contact size. Forming pulses for the film grown in 10 mTorr measured using 0.3 µm and 6.5 µm contacts are shown in Fig. 6(b). The data show that larger diameter top contacts require a smaller voltage during the electroforming step, in agreement with Ref. 12.

We also compared the threshold switching behavior of these films. Figure 7(a) shows a circuit diagram for pulsed IV measurements. Triangular voltage pulses with different pulse widths ($\tau$) were applied across the films and the output current was measured across a test resistor $R_s$. A non-linear current switching characteristic was observed for pulse widths $\tau \leq 10$ µs. Output pulses for both samples for $\tau \leq 10$ µs are plotted in Fig. 7(b) and 7(d). The device from the film grown in 1 mTorr had a smaller turn-on voltage $V_{ON}$ = 1.5 V and a smaller current ON/OFF ratio (~5) than the one for the device from the film grown in 10 mTorr pressure, $V_{ON}$ = 2 V and a ratio ~10. This can be explained by the more insulating character of the film grown in 10 mTorr. In addition, the device made from the film grown in 10 mTorr had more abrupt ON- and OFF-switching. An enlarged view of an output current pulse measured on the film grown in 10 mTorr growth pressure is shown in Fig. 7(c). The typical OFF-to-ON switching time ($\tau_{ON}$) was found to be 25 ns, which is fast enough for a current selection element in many RRAM device prototypes.[33] This result held true for 4 other samples grown using similar growth conditions. ON/OFF switching times for the film grown in 1 mTorr [Fig. 7(b)] were approximately twice as long as the ON/OFF times measured for the films grown in 10 mTorr pressure. Here, the switching time is considered only for extremely sharp region [shaded region in Fig. 7(c)]. Thus, the device from the film with lower $NbO_2$ content (grown in 10 mTorr pressure) can provide the better performance if used in a current switching element. Interestingly, a current overshoot was observed in case of the film grown in 10 mTorr growth pressure [Fig. 7(c) and (d)], which was not observed for one grown in 1 mTorr [Fig. 7(b)]. The current overshoot is likely caused by the switching time $\tau_{ON}$ being shorter than the associated capacitance ($C_p$) discharge time ($\sim R_{ON}C_p$).[34] The fact that no overshoot seems to happen for films grown in 1 mTorr pressure is further proof of a slower switching behavior.

We also note that films grown in pressures greater than 10 mTorr had a significant amount of the insulating $Nb_2O_5$ phase and no threshold switching behavior was observed.

### E. Self-sustained current oscillations

Finally, we discuss self-sustained current oscillations in our vertical devices. Figure 8(a) shows the connection diagram. The sample was connected in series to a load resistor ($R_L$) of 10 kΩ and a sensing resistor ($R_S$) of 51 Ω; the latter was used to measure the current. The sample is shown as a switchable resistor connected to a parallel capacitor, representing the intrinsic capacitance of the insulating (OFF) state. A low frequency input voltage ($V_{in}$) with a triangular or rectangular wave shape was applied and the output signal (i.e. current, $I_{out}$) was measured across $R_S$.

Figure 8(b) shows a current response after a triangular voltage pulse was applied. Current oscillations were observed only within the range of the input voltage 2 V < $V_{in}$ < 5 V [red, dotted line in zoomed view of Fig. 8(b)]. These oscillations are a result of a thermally induced negative differential resistance region.[4,35,36] After the film material temperature increases due to the Joule heating, the current keeps on rising even while the input voltage is decreasing. The threshold current is highly unstable in this region, so that by choosing appropriate circuit parameters, such as $R_L$ and $V_{in}$, self-sustained current oscillations can be induced.[37]

Figure 8(c) shows the output current characteristics of the film grown in 1 mTorr and measured by applying a rectangular pulse with ~ 9 µs pulse width and 4.4 V in amplitude. Before applying $V_{in}$, the $NbO_2$ device was in its initial high resistance state, so the associated capacitor was charged when the voltage was applied. The increase in current caused it to switch to the low resistance state, resulting in a discharge of the capacitor. During this step, the largest voltage $V_{in}$ drop occurred across $R_L$ and so the film turned itself insulating when the sample voltage was reduced. A rapid repetition of the process provided oscillations in the output current.

Highly stable oscillations were obtained during positive and negative cycles of the input voltage. Oscillations lasted over 48 hours ($10^{12}$ cycles) with 5% fluctuation in frequency and 33% fluctuation in amplitude, which is the most stable oscillator so far reported for $NbO_2$. The peak-to-peak amplitude ($I_{pp}$) of current oscillations was about 1.1 mA. A Fast Fourier Transform (FFT) of the signal showed that the first order frequency component was at 11 MHz [Fig. 8(c)] for an input voltage of 4.5 V.

We studied the dependence of the oscillation frequency on input voltage in films grown at 1 and 10 mTorr total growth pressures, i.e. in films with different $NbO_2$-content (Fig. 9). A closer view of oscillations is shown in the inset to Fig. 9 with similar frequencies of ~ 11 MHz for both films. Aside from the main frequency oscillations, other harmonics were present.

For the film grown in 1 mTorr, the range of $V_{in}$ where oscillations were observed was found to be between 2.5 V to 4.8 V, in agreement with the study on the same film under triangular pulse application [Fig. 8(b)]. The oscillation frequency $f_{osc}$ increased upon increasing input voltage amplitude $V_{in}$ until $V_{in}$ = 4 V. Upon further increase of $V_{in}$, $f_{osc}$ started to decrease and oscillations disappeared. Lalevic et al. have reported a similar trend in frequency variation: linear increment followed by a saturated region using polycrystalline thin film devices.[38] For the film grown in 10 mTorr, the frequency increased linearly with increasing $V_{in}$ and no saturation region existed. The range of $V_{in}$ resulting in stable oscillations for the film was higher, between 4 V and 7.5 V. Because of the lower $NbO_2$ content, the film grown in 10 mTorr pressure had higher threshold voltage and thus a larger range of $V_{in}$ which produced oscillations.

## IV. Conclusions

In conclusion, we studied structural and electric properties of $NbO_2$ films grown on TiN coated $SiO_2$/Si substrate using different growth pressures. θ-2θ and grazing incidence x-ray diffractometry verified the polycrystalline growth of $NbO_2$. Threshold switching voltage ($V_{th}$) showed thickness dependence and off-state leakage current was described by the Poole-Frenkel conduction mechanism. The threshold switching behavior of the device depended on the contact size, with a dependence on the diameter of the contact for small contacts and on the area for large contacts. Pulsed *I-V* measurements performed on the sample grown in 10 mTorr pressure showed a current switching time of ~25 ns. Highly stable self-oscillatory current behavior with oscillation frequencies above 5 MHz were demonstrated. Higher $Nb_2O_5$ content resulted in self-sustained oscillations of higher frequencies.

## Acknowledgements

This work was supported in part by FAME, one of six centers of STARnet, a Semiconductor Research Corporation program sponsored by MARCO and DARPA (Contract # 2013-MA-2382) and the West Virginia University Shared Research Facilities

**References:**

[1] Morin, F. J. Phys. Rev. Lett. 3, 34 (1959)

[2] M. Son, J. Lee, J. Park, J. Shin, G. Choi, S. Jung, W. Lee, S. Kim, S. Park, and H. Hwang, IEEE Electron Device Letters **32**, 1579 (2011)

[3] A. Beaumont, J. Leroy, J.-C. Orlianges, and A. Crunteanu, J. Appl. Phys. **115**, 154502 (2014)

[4] M. D. Pickett, G. Medeiros-Ribeiro and R. S. Williams, *Nat. Mater.* **12,** 114 (2013)

[5] S. Li, X. Liu, S. K. Nandi, D. K. Venkatachalam and R. G. Elliman, Nanotechnology 28, 125201 (2017)

[6] S. Kim, J. Park, J. Woo, C. Cho, W. Lee, J. Shin, G. Choi, S. Park, D. Lee, B. H. Lee, H. Hwang, *Microel. Eng.* **107,** 33 (2013)

[7] X. Liu, S. K. Nandi, D. K. Venkatachalam, K. Belay, S. Song, and R. G. Elliman, IEEE Electron Device Letters 35, 1055 (2014)

[8] H. R. Phillipp, and L. M. Levinson, *J. Appl. Phys.* **50** 4814 (1979)

[9] J. C. Lee, W. W, Durand, *J. Appl. Phys* **56** 3350 (1984)

[10] S. H. Shin, T. Halpern, and P. M. Raccah, *J. Appl. Phys.* 48, 3150 (1977)

[11] X. Liu, S. M. Sadaf, M. Son, J. Shin, J. Park, J. Lee, S. Park, and H. Hwang, Nanotechnology 22, 472702 (2011)

[12] E. Cha, J. Park, J. Woo, D. Lee, A. Prakash, and H. Hwang, Appl. Phys. Lett. **108**, 153502 (2016)

[13] S. Li, X. Liu, S. K. Nandi, D. K. Venkatachalam, and R. G. Elliman, Appl. Phys. Lett. 106, 212902 (2015)

[14] E. Cha, J. Woo, D. Lee, S. Lee, J. Song, Y. Koo, J. Lee, C.G. Park, M.Y. Yang, K. Kamiya, K. Shiraishi, B. Magyari-Kope, Y. Nishi, and H. Hwang, in 2013 IEEE Int. Electron Devices Meet. IEEE, 2013, pp. 10.5.1

[15] A. A. Bolzan, C. Fong, and B. J. Kennedy, *J. Solid St. Chem.* **113**, 9 (1991)

[16] K. Sakata, I. Nishida, M. Matsushima, and T. Sakata, *J. Phys. Soc. Japan* **27,** 506 (1969)

[17] D. J. Wouters, R. Waser, and M. Wuttig, Proceedings of the IEE, 103, 1274 (2015)

[18] J.H. Lee, E.J. Cha, Y.T. Kim, B.K. Chae, J.J. Kim, S.Y. Lee, H.S. Hwang, C.G. Park, Micron **79**, 101 (2015)

[19] T. Joshi, T. R. Senty, P. Borisov, A. D. Bristow and D. Lederman, J. Phys. D: Appl. Phys. **48**, 335308 (2015)

[20] T. Joshi, P. Borisov, and D. Lederman, AIP Advances **6**, 125006 (2016)

[21] Grier, D., McCarthy, G., North Dakota State University, Fargo, North Dakota, USA. ICDD Grant-in-Aid (1991)

[22] M. Bj¨orck and G. Andersson, Journal of Applied Crystallography **40**, 1174 (2007)

[23] C. Lin, A. Posadas, T. Hadamek, and A. A. Demkov, PHYSICAL REVIEW B **92**, 035110 (2015)

[24] F. J. Wong, N. Hong, and S. Ramanathan, Phys. Rev. B **90**, 115135 (2014)

[25] I. H. Inoue, S. Yasuda, H. Akinaga, and H. Takagi, Phys. Rev. B 77, 035105 (2008)

[26]R. Meyer, L. Schloss, J. Brewer, R. Lambertson, W. Kinney, J. Sanchez, and D. Rinerson, Proc. 9th annual non-volatile Memory Tech. Symp. pp 54 (2008)

[27] J. Frenkel , Phys. Rev. 54, 647 (1938)

[28]A. Ongaro, A. Pillonnet, IEEE Proceeding 138, 127 (1991)

[29]S. Slesazeck, H. Mahne, H. Wylezich, A. Wachowiak, J. Radhakrishnan, A. Ascoli, R. Tetzlaff and T. Mikolajicka, RSC Adv. 5, 102318 (2015)

[30]G. A. Gibson, S. Musunuru, J. Zhang, K. Vandenberghe, J. Lee, C.-C. Hsieh, W. Jackson, Y. Jeon, D. Henze, Z. Li, and R. S. Williams, Appl. Phys. Lett. **108**, 023505 (2016)

[31]C. Funck , S. Menzel , N. Aslam , H. Zhang , A. Hardtdegen, R. Waser, and S. Hoffmann-Eifert, Adv. Electron. Mater. 2, 1600169 (2016)

[32]K. M. Kim, J. Zhang, C. Graves, J. J. Yang, B. J. Choi, C. S. Hwang, Z. Li, and R. S. Williams, Nano Letters 16, 6724 (2016)

[33]S.-E. Ahn, M.-J. Lee, Y. Park, B. S. Kang, C. B. Lee, K. H. Kim, S. Seo, D.-S. Suh, D.-C. Kim, J. Hur, W. Xianyu, G. Stefanovich, H. A. Yin, I.- K. Yoo, A.-H. Lee, J.-B. Park, I.-G. Baek, and B. H. Park, Adv. Mater. 20, 924 (2008)

[34] P. R. Shrestha, D. M. Nminibapiel, J. P. Campbell, J. T. Ryan, D. Veksler, H. Baumgart and K. P. Cheung, IEEE Trans. Electron. Dev. 65, 108 (2018)

[35]Chopra K L 1963 Proc. IEEE 51 941

[36]Geppert D V 1963 Proc. IEEE 51 223

[37]S. O. Pearson and H. St. G. Anson, Proc. Phys. Soc. of London 34, 204 (1921)

[37]B. lalevic and M. Shoga, Thin solid films, 75, 199 (1981)

**List of Figures:**

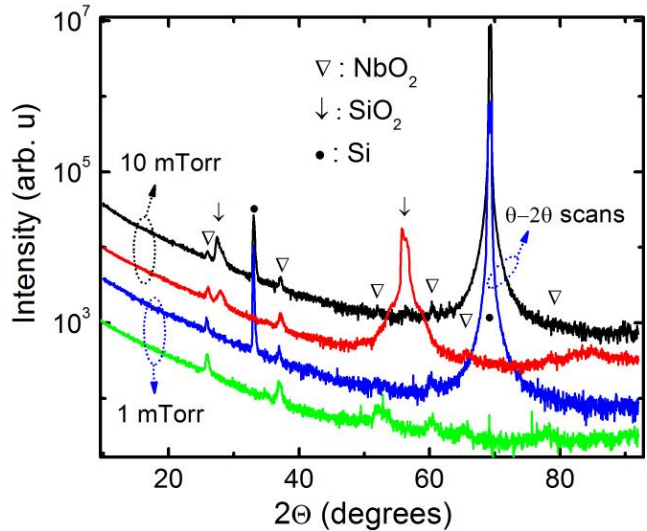

Figure 1: X-ray diffraction spectra. θ-2θ (black and blue curves) and grazing incidence x-ray diffraction (GIXRD) (red and green curves) patterns of NbO$_2$ films grown in 1 (green and blue curves) and 10 mTorr (red and black curves) total pressure. Sharp peaks denoted by '●' and '↓' are from (100) Si and (010) SiO$_2$, respectively. The grazing incidence angle was $\alpha = 3^0$.

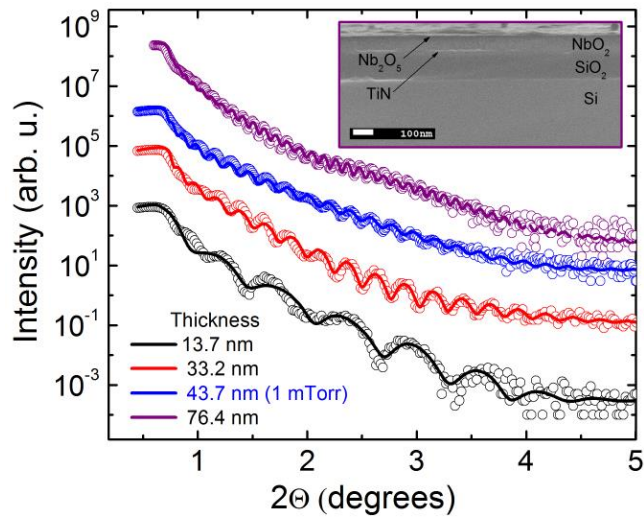

Figure 2: X-ray reflectivity (XRR) data (open circles) for NbO$_2$ films with the corresponding fits (solid lines). All films were grown in 10 mTorr total pressure, except the 43.7 nm film grown in 1mTorr total pressure. Inset figure shows the cross-sectional SEM image of the 76.4 nm thin film.

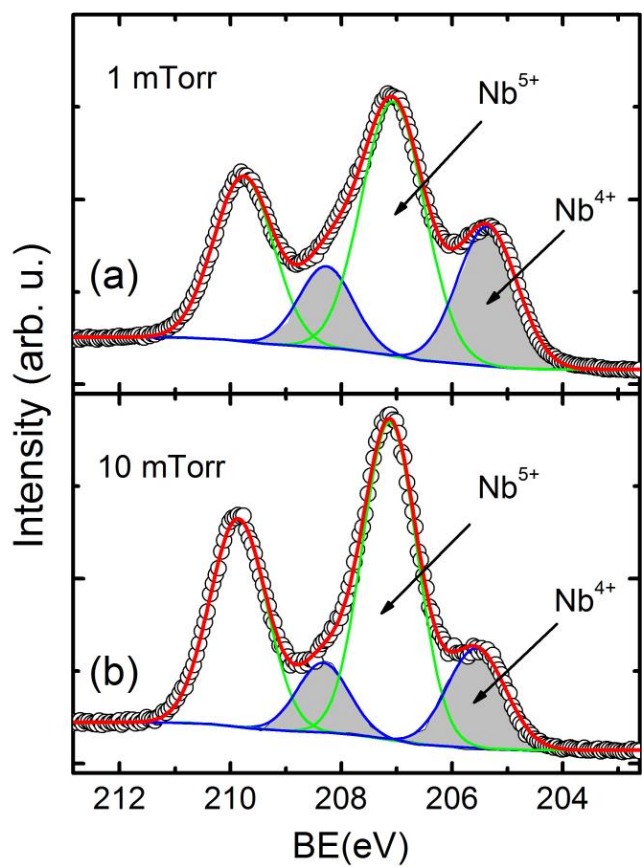

Figure 3: X-ray photoelectron spectroscopy (XPS) data for Nb $3d_{5/2}$ and $Nb_{3/2}$ spectra of films grown in 1 mTorr (a) and 10 mTorr (b) growth pressures with deconvoluted peak fits. The shaded areas are the contribution from the $Nb^{4+}$ valency state. The molar content of $NbO_2/Nb_2O_5$, estimated using areas covered by $3d_{5/2}$ level speaks, was found as 34.2%/65.8% and 25.1%/74.9% for the film grown in 1 and 10 mTorr, respectively.

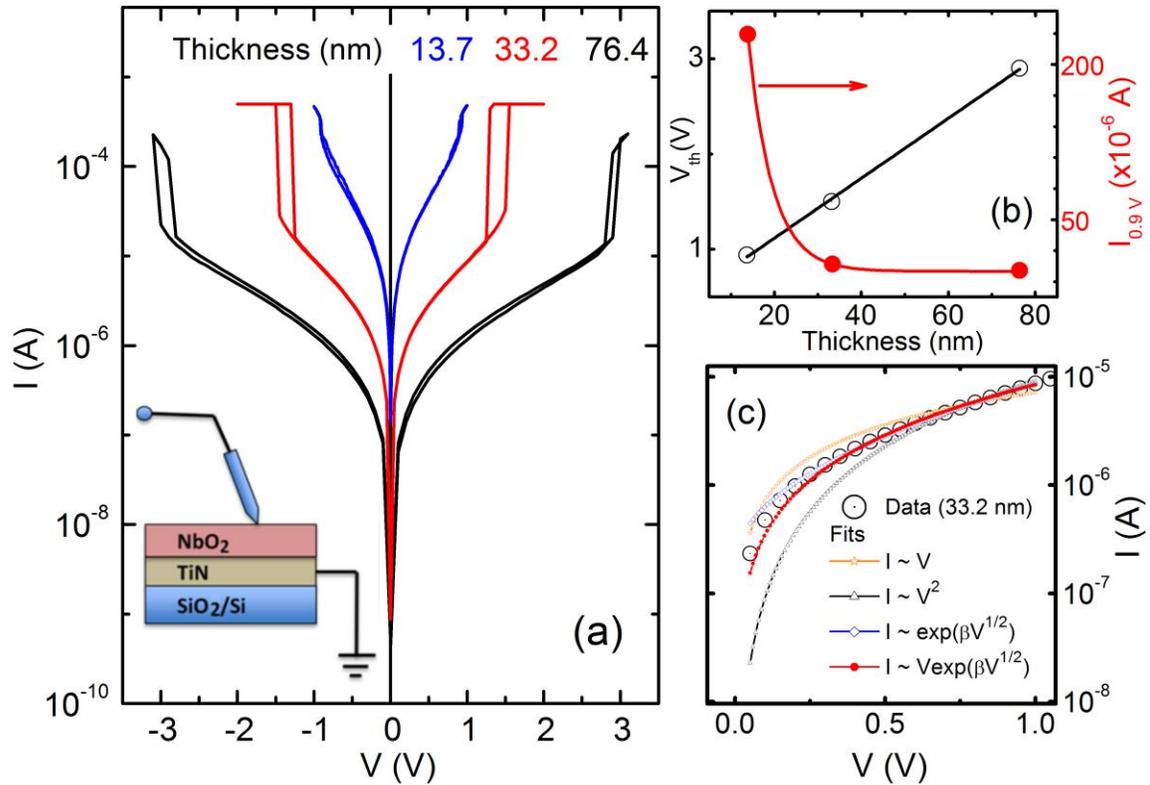

Figure 4: (a) *I-V* curves measured using probe station tungsten tip at RT in air on films of different thicknesses. Inset: schematics of sample structure and measurement setup. (b) Threshold voltage ($V_{th}$) and current measured at V = 0.9 V plotted against film thickness. Solid lines are exponential and linear fits, respectively. (c) Experimental *I-V* curve data for 33.2 nm NbO$_2$ film below 1 V plotted together with fits to the following conduction models: Ohmic ($I \sim V$), space-charge-limited ($I \sim V^2$), Schottky [$I \sim \exp(\beta V^{\frac{1}{2}})$] and Poole-Frankel conduction [$I \sim V\exp(\beta V^{1/2})$].

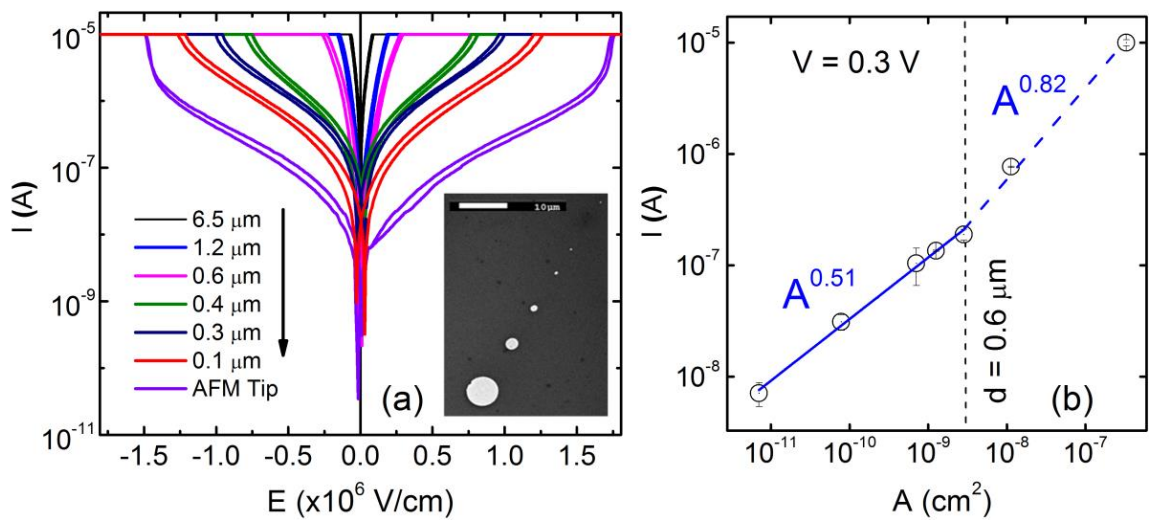

Figure 5: Contact size dependent *I-V*. (a) *I-V* curves measured using conducting probe AFM at RT in air for film grown in 10 mTorr pressure using different electrode sizes (inset: SEM image of the top electrodes). (b) Current measured at 0.3 V vs. top contact area A. Solid lines represent linear fits on log-log scale to the current I∝ $A^n$, where A is the area of the top electrode, and n is the exponent.

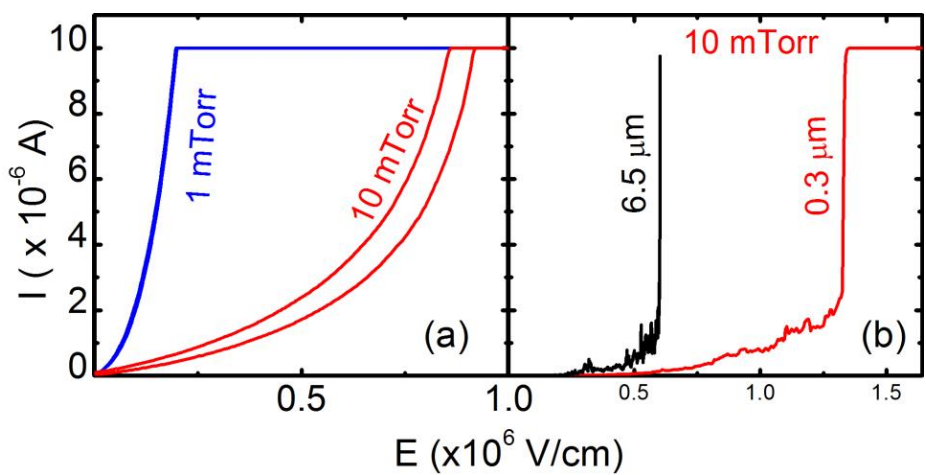

Figure 6: (a) *I-V* curves obtained from the films grown in 1 mTorr and 10 mTorr films using 0.3 µm contacts. (b) Two of the forming pulses for the film grown in 10 mTorr with contacts 6.5 and 0.3 µm.

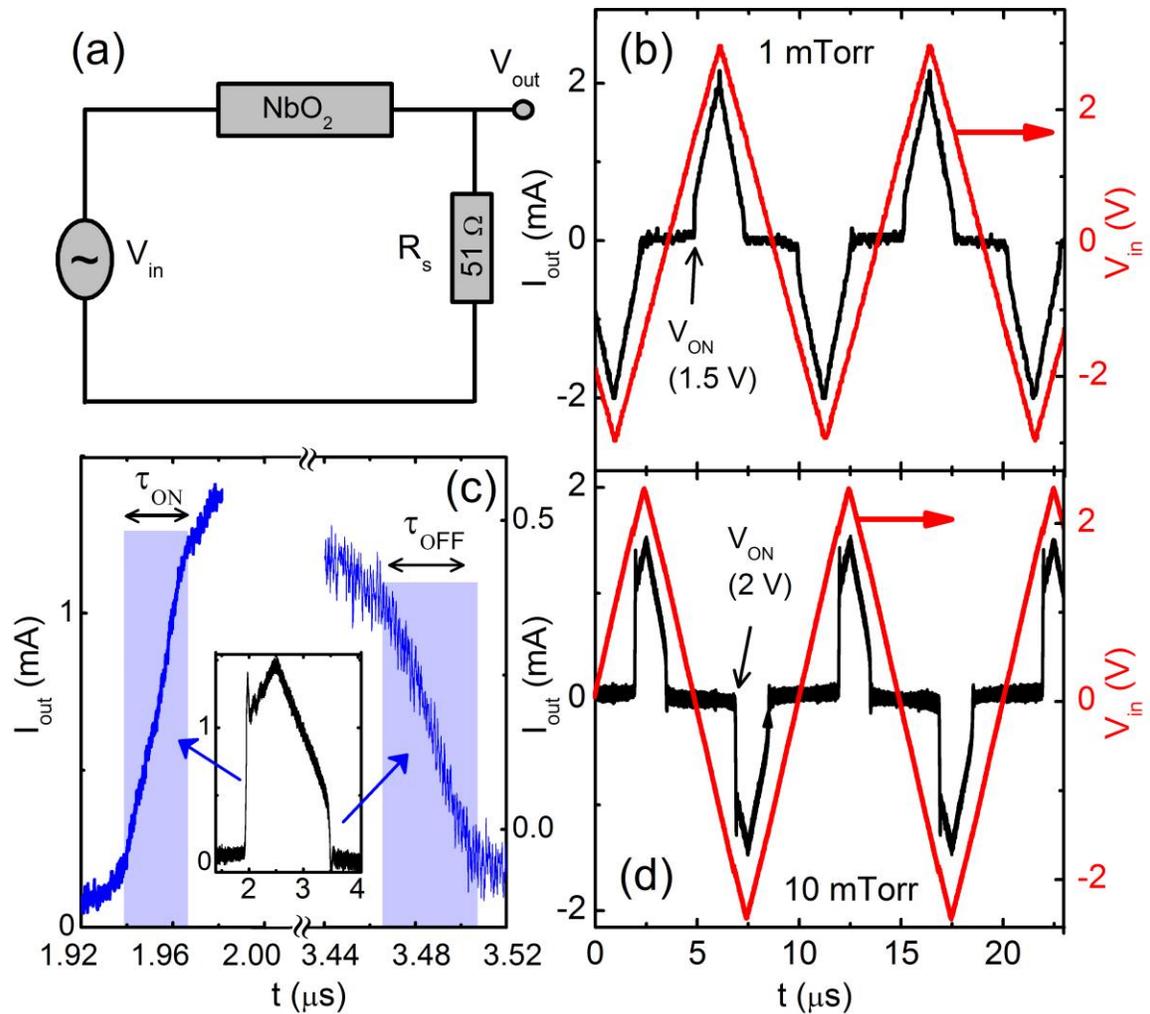

Figure 7: Pulsed *I-V* measurement. (a) Circuit diagram. (b), (d) Pulsed input (right scale) and output current characteristics obtained from films grown in 1 and 10 mTorr films, respectively. (c) A zoomed part of (d) showing current switching times: $\tau_{ON}$ = 25 ns and $\tau_{OFF}$ = 40 ns .

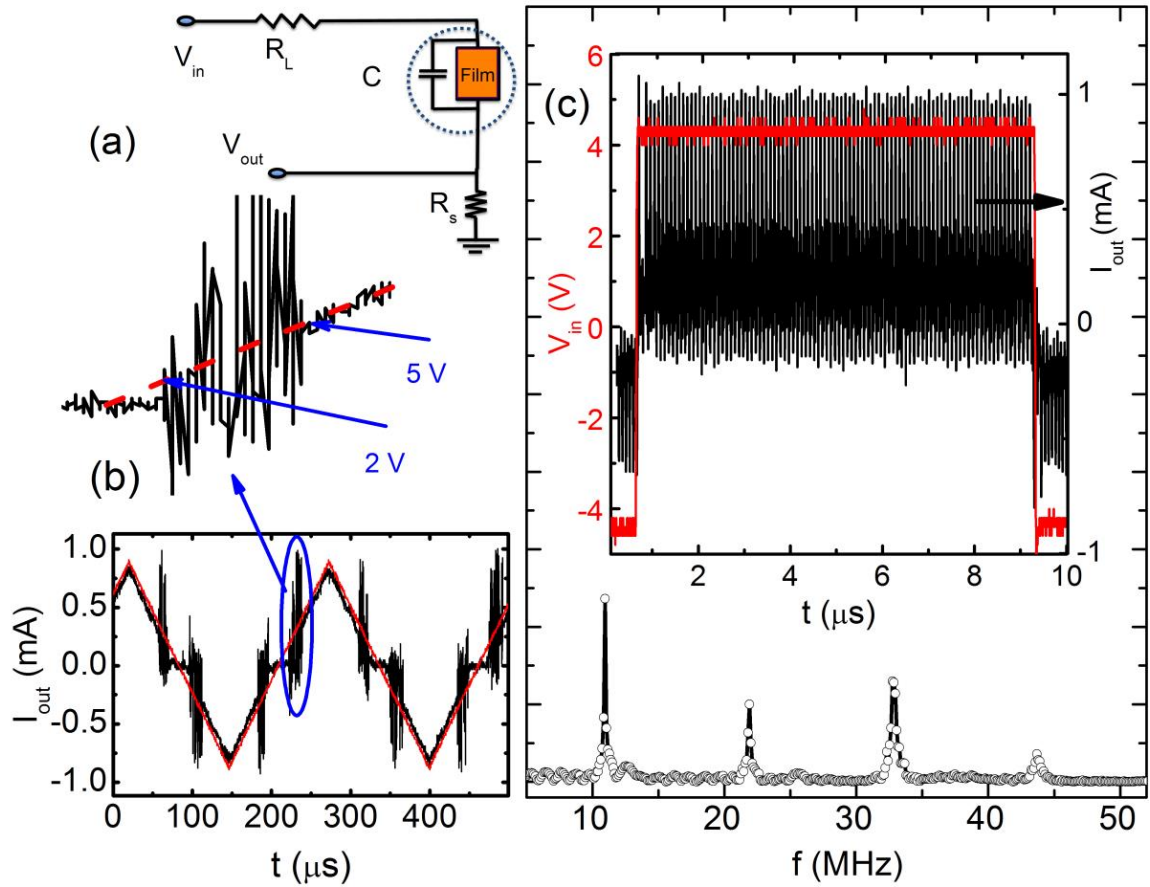

Figure 8: Self-sustained current oscillations obtained from the film grown in 1 mTorr pressure. (a) Schematic diagram for a current oscillator. (b) Current oscillations produced by a triangular pulse field. Zoomed figure showed the range of input voltage contributing for oscillation was 2 V to 5 V. (c) Self-sustained current oscillations measured from the film grown in 1 mTorr growth pressure using a rectangular pulse field with Fast Fourier Transform (FFT) showing frequency components.

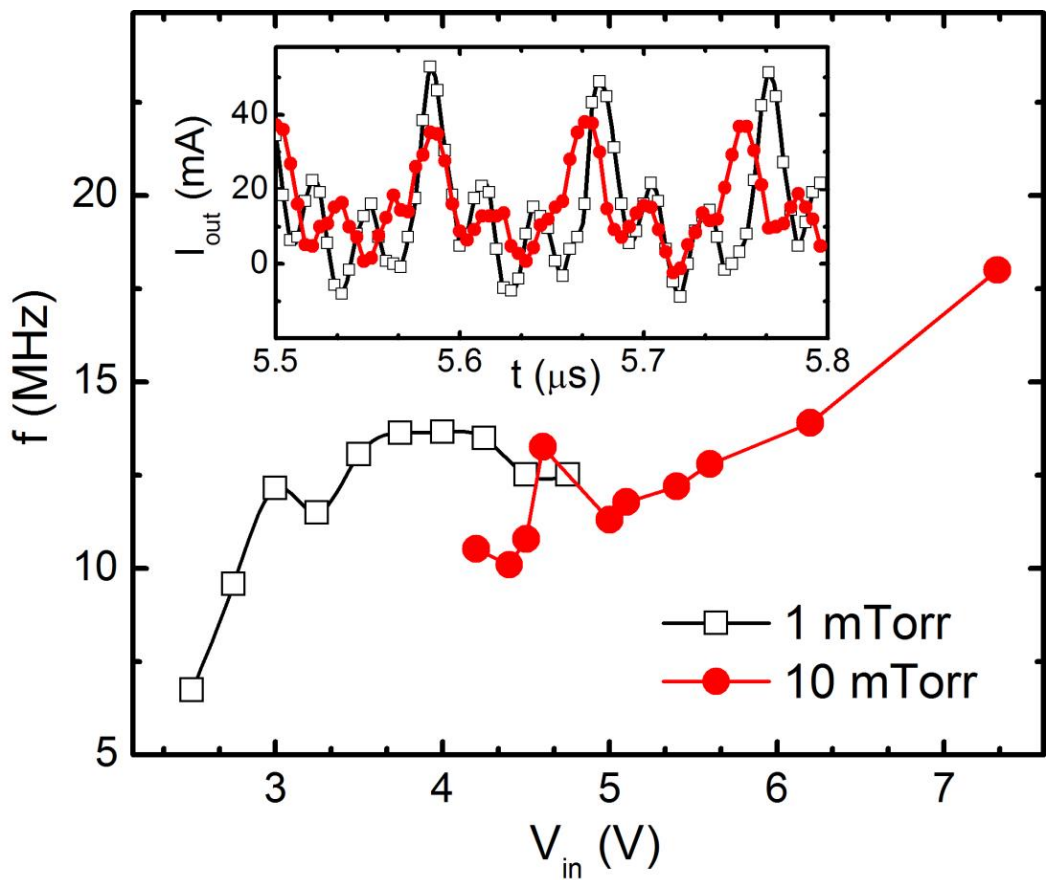

Figure 9: Input *V* versus *f* of films grown in 1 and 10 mTorr pressures. Inset figure shows zoomed view of current oscillations with similar frequency about 11 MHz measured on both films grown in 1 and 10 mTorr growth pressure.